\documentclass[aps,prb,10pt,amsmath,amssymb,footinbib,showpacs,twocolumn]{revtex4-2}
\usepackage{amsmath}
\usepackage{amssymb}
\usepackage{amsthm}
\usepackage{setspace}
\usepackage{graphicx}
\usepackage{braket}
\usepackage{mathrsfs}
\usepackage{float}
\usepackage[colorlinks = true,linkcolor = blue,urlcolor  = blue,citecolor = blue,anchorcolor = blue]{hyperref}
\usepackage[utf8]{inputenc}
\usepackage[english]{babel}
\usepackage{bm}
\usepackage{csquotes}
\usepackage{comment}
\begin{document}

\title{Nonlinear anomalous Hall effect in three-dimensional chiral fermions}
\author{Azaz Ahmad}
\affiliation{School of Physical Sciences, Indian Institute of Technology Mandi, Mandi 175005, India.}
\author{Gautham Varma K.}
\affiliation{School of Physical Sciences, Indian Institute of Technology Mandi, Mandi 175005, India.}
\author{Gargee Sharma}
\affiliation{School of Physical Sciences, Indian Institute of Technology Mandi, Mandi 175005, India.} 

\begin{abstract}
Chiral fermionic quasiparticles emerge in certain quantum condensed matter systems such as Weyl semimetals, topological insulators, and spin-orbit coupled noncentrosymmetric metals.  Here, a comprehensive theory of the chiral anomaly-induced nonlinear anomalous Hall effect (CNLAHE) is developed for three-dimensional chiral quasiparticles, advancing previous models by rigorously including momentum-dependent chirality-preserving and chirality-breaking scattering processes and global charge conservation. Focusing on two specific systems--Weyl semimetals (WSMs) and spin-orbit coupled non-centrosymmetric metals (SOC-NCMs), we uncover that the nonlinear anomalous Hall conductivity in WSMs shows nonmonotonic behavior with the Weyl cone tilt and experiences a `strong-sign-reversal' with increasing internode scattering, diverging from earlier predictions. For SOC-NCMs, where nonlinear anomalous Hall conductivity has been less explored, we reveal that unlike WSM, the orbital magnetic moment alone can drive a large CNLAHE with distinctive features: the CNLAH conductivity remains consistently negative regardless of interband scattering intensity and exhibits a quadratic dependence on the magnetic field, contrasting the linear dependence in WSMs. Furthermore, we discover that in SOC-NCMs the Zeeman coupling of the magnetic field acts like an effective tilt term which can further enhance the CNLAH current.
These findings offer fresh insights into the nonlinear transport dynamics of chiral quasiparticles and can be verified in upcoming experiments on such materials. 
\end{abstract}

\maketitle

\section{Introduction}
The concept of chiral particles originates from high-energy physics~\cite{peskin1995introduction}. 
While electrons, protons, and neutrons have chiral aspects in their interactions and internal structure, they are not fundamentally chiral particles because of their finite mass. On the other hand, the existence of massless chiral fermions is now well-established in condensed matter systems~\cite{hasan2021weyl}. They emerge in certain materials as quasiparticles exhibiting behavior analogous to the theorized chiral fermions in particle physics. Two prominent examples of these materials are topological insulators (TIs)~\cite{hasan2010colloquium,qi2011topological} and Weyl semimetals (WSMs)~\cite{hosur2013recent,armitage2018weyl}. TIs have gapped bulk states, while their boundary states are massless and chiral. 
In contrast, WSMs have gapless bulk chiral states (Weyl fermions) that are topologically protected by a non-vanishing Chern number, which is equivalent to the chirality quantum number. 
Nielsen \& Ninomiya, who first studied the regularization of Weyl fermions on a lattice, showed that they must occur in pairs of opposite chiralities~\cite{nielsen1981no,nielsen1983adler}, thus leading to the conservation of both chiral charge and global charge in absence of any gauge fields. Probing the chirality of the emergent Weyl fermions in WSMs has been of utmost theoretical and experimental interest since the past decade~\cite{Yan_2017,hasan2017discovery,burkov2018weyl,ong2021experimental,nagaosa2020transport,lv2021experimental,ahmad2024geometry}. 

In the presence of external electromagnetic fields, the chiral charge is not conserved, which is the celebrated chiral anomaly (CA) or the Adler-Bell-Jackiw (ABJ) anomaly~\cite{adler1969axial} of Weyl fermions. The non-conservation of chiral charges leads to an anomaly-induced current that may be verified in WSMs by measuring its transport and optical properties~\cite{parameswaran2014probing,hosur2015tunable,goswami2015optical,goswami2013axionic,son2013chiral,burkov2011weyl,burkov2014anomalous,lundgren2014thermoelectric,sharma2016nernst,kim2014boltzmann,zyuzin2017magnetotransport,cortijo2016linear}. Interestingly, CA has been proposed to occur in systems that are not WSMs~\cite{gao2017intrinsic,dai2017negative,andreev2018longitudinal,wang2018intrinsic,nandy2018berry,fu2020quantum,pal2021berry,wang2021helical,sadhukhan2023effect,PhysRevB.105.L180303,das2023chiral,varma2024magnetotransport}. The quasiparticles, in this case, are not necessarily massless but have a notion of chirality due to their underlying spinor structure.
This has led to the generalization that CA may manifest in any system with nonzero Berry flux through the Fermi surface, irrespective of the energy dispersion, number of Weyl nodes, or the underlying symmetries of the Hamiltonian~\cite{PhysRevB.105.L180303}. A specific example is that of spin-orbit-coupled (SOC) non-centrosymmetric metals (NCMs) that host nonrelativistic fermions but have multiple Fermi surfaces with fluxes of opposite Berry curvature~\cite{PhysRevB.105.L180303}. A few recent studies have investigated CA-induced electronic and thermal transport in SOC-NCMs~\cite{verma2019thermoelectric,PhysRevB.105.L180303,das2023chiral,varma2024magnetotransport}.
While some band properties in SOC-NCMs may be similar to those of WSMs, their transport responses are strikingly different~\cite{varma2024magnetotransport}.

The transport of chiral quasiparticles in condensed matter systems is affected by two key scattering processes: (i) chirality-breaking, and (ii) chirality-preserving scattering. In the context of WSMs, these are also known as (i) internode scattering (chirality-breaking), and (ii) intranode scattering (chirality-preserving), respectively, as Weyl fermions of opposite chiralities live at different Weyl nodes (or valley points) in the momentum space. Since SOC-NCMs have just one relevant nodal point, but with multiple Fermi surfaces with opposing fluxes of the Berry curvature, the two types of scattering mechanisms refer to  (i) interband (chirality-breaking) and (ii) intraband (chirality-preserving) scattering, respectively.  
Notably, the chiral anomaly manifests by the first process--the chirality-breaking scattering, which is governed by corresponding scattering timescale $\tau_\mathrm{inter}$. The second process that preserves chirality, which is not directly related to the anomaly, is governed by a timescale $\tau_\mathrm{intra}$. Nevertheles, a series of earlier works~\cite{kim2014boltzmann,lundgren2014thermoelectric,cortijo2016linear,sharma2016nernst,zyuzin2017magnetotransport,das2019berry,kundu2020magnetotransport} have primarily focused on the role of $\tau_\mathrm{intra}$ while investigating CA-induced transport, while neglecting $\tau_\mathrm{inter}$. This is sometimes justified by stating that the chirality preserving scattering often dominates, i.e., $1/\tau_\mathrm{inter}\ll 1/\tau_\mathrm{intra}$. 
But even in the approximation that $\tau_\mathrm{inter}\gg\tau_\mathrm{intra}$, the analysis in most of the previous studies is flawed for two main reasons: (i) they neglect global charge conservation, and (ii) they assume a momentum-independent scattering time, which was recently shown to be inaccurate for chiral quasiparticles~\cite{sharma2023decoupling}. 
Recent studies have refined the understanding and analysis of transport in chiral Weyl fermions by moving beyond previous assumptions, leading to some striking and significant predictions in linear magnetotransport~\cite{knoll2020negative,sharma2020sign,sharma2023decoupling,ahmad2021longitudinal,ahmad2023longitudinal}.

Apart from inducing currents proportional to the applied field (linear response), chirality-violating processes can induce nonlinear effects as well, such as the nonlinear Hall effect~\cite{PhysRevB.103.045105}. In an inversion symmetry-broken Weyl semimetal with tilted Weyl cones, a nonlinear Hall effect can be induced by the chiral anomaly, known as the chiral anomaly-induced nonlinear anomalous Hall effect (CNLAHE), which is the combined effect of the Berry curvature-induced anomalous velocity $\mathbf{v}_{\mathrm{anom}}=({e}/{\hbar})\mathbf{E}\times \mathbf{\Omega_{k}}$~\cite{xiao2010berry}  and the chiral anomaly. The effect is nonzero when the Fermi surface is asymmetric and the Hamiltonian exhibits broken inversion symmetry. In WSMs, the tilt of the Weyl cone creates an asymmetric Fermi surface around the projection of the Weyl node on the Fermi surface. It is important to note that the chiral anomaly-induced nonlinear Hall effect (CNLAHE) is distinct from the CNLHE caused by the Berry curvature dipole (BCD) \cite{sodemann2015quantum}, as the latter can occur even without an external magnetic field.
Previous works on CNLAHE~\cite{li2021nonlinear, nandy2021chiral,zeng2022chiral} assume that the internode scattering rate is much lower than the intranode scattering rate, or in other words $\tau_\mathrm{inter}\gg\tau_\mathrm{intra}$, thereby neglecting the role of internode scattering. Furthermore, the analysis suffers from the aforementioned shortcomings: (i) neglecting global charge conservation, and (ii) assumption of a momentum-independent scattering time, both of which breakdown for chiral quasiparticles of multiple flavors.

In this work, we present a complete theory of the nonlinear anomalous Hall effect, correctly including the effects of chirality-breaking and chirality-preserving scattering, retaining their full momentum dependence, and incorporating global charge conservation. Our theory is generic and works for any system with chiral quasiparticles of multiple flavors, however, we focus on two particular systems of experimental interest: (i) Weyl semimetal, and (ii) spin-orbit coupled noncentrosymmetric metal. We find that in Weyl semimetals the nonlinear anomalous Hall conductivity is a nonmonotonic function of the Weyl cone tilt, which is in contrast to earlier studies~\cite{li2021nonlinear,nandy2021chiral,zeng2022chiral}. Furthermore, we also find that sufficiently strong internode scattering (not considered in earlier works) flips the sign of conductivity leading to `strong-sign-reversal'. Additionally, we also examine the effect of strain (also not considered in prior works), and find that strain-induced chiral gauge field also gives rise to nonlinear anomalous Hall effect but without any `strong-sign-reversal'. 
The nonlinear anomalous Hall conductivity has not been earlier analyzed in spin-orbit coupled non-centrosymmetric metals, which forms another important focus of this work. While, the chiralities of quasiparticles in WSMs and SOC-NCMs is exactly the same, their nonlinear current response is remarkably distinct from each other. 
Interestingly, we discover that unlike WSMs, the anomalous orbital magnetic moment in SOC-NCMs can drive a large CNLAH current when the electric and magnetic fields are noncollinear. We further find that including the effect of Zeeman coupling of the magnetic field acts like an effective tilt term which tilts the Fermi surfaces of both the chiral flavors in the same direction, thereby further enhancing the CNLAH current. We highlight significant differences between the nonlinear conductivity obtained for WSMs and SOC-NCMs. First, CNLAHE can be driven in SOC-NCMs by anomalous orbital magnetic moment, unlike WSMs where the cones must be necessarily tilted. Second, CNLAH conductivity in WSMs flips its sign with sufficiently strong internode scattering, unlike SOC-NCMs where CNLAH conductivity remains always negative even for sufficiently high interband scattering (although both these processes break the quasiparticle chirality). Third, CNLAH conductivity is linear in $B$ for WSMs but is quadratic in $B$ for SOC-NCMs. Lastly, the angular dependence of CNLAH conductivity is strikingly different from that of WSMs. 

In Section II, we present the Boltzmann theory where an analytical ansatz to the electron distribution function is derived. Secion III and IV discuss the CNLAH conductivity is WSMs and SOC-NMCs, respectively. We conclude in Section V. 

\section{Maxwell-Boltzmann transport theory}
\label{Sec:Maxwell-Boltzmann transport theory}
We use the semiclassical Maxwell-Boltzmann formalism to describe the dynamics of three-dimensional chiral fermions in the presence of external electric and magnetic fields. The non-equilibrium distribution function $f^{\chi}_{\mathbf{k}}$ describing fermions with chirality $\chi$, evolves as:
\begin{align}
\dfrac{\partial f^{\chi}_{\mathbf{k}}}{\partial t}+ {\Dot{\mathbf{r^{\chi}_{\mathbf{k}}}}}\cdot \mathbf{\nabla_r}{f^{\chi}_{\mathbf{k}}}+\Dot{\mathbf{k^{\chi}}}\cdot \mathbf{\nabla_k}{f^{\chi}_{\mathbf{k}}}=I_{coll}[f^{\chi}_{\mathbf{k}}],
\label{MB_equation}
\end{align}
with $f^{\chi}_\mathbf{k} = f_{0} + g^{\chi}_{\mathbf{k}}$ + $h^{\chi}_{\mathbf{k}}$, where $f_{0}$ is standard Fermi-Dirac distribution, and $g^{\chi}_{\mathbf{k}}$ and $h^{\chi}_{\mathbf{k}}$ are deviations up to the first and second order in electric field ($E$), respectively. Without loss of generality, we fix the electric field along the $z-$direction and express the deviations as:
\begin{align}
g^{\chi}_\mathbf{k}&= -e\left({\dfrac{\partial f_{0}}{\partial {\epsilon}}}\right){\Lambda^{\chi}_\mathbf{k}} E\nonumber\\
h^{\chi}_\mathbf{k}&= -e\left({\dfrac{\partial g^{\chi}_\mathbf{k}}{\partial {\epsilon}}}\right){\Gamma^{\chi}_\mathbf{k}} E\nonumber\\
&=e^2\left(\left(\frac{\partial^2 f_0}{\partial \epsilon^2}\right)\Lambda^\chi_\mathbf{k}+\left(\frac{\partial \Lambda^\chi_\mathbf{k}}{\partial \epsilon}\right)\left(\frac{\partial f_0}{\partial \epsilon}\right)\right) \Gamma^{\chi}_\mathbf{k} E^2
\label{Eq:g1},
\end{align}
where ${\Lambda}^{\chi}_\mathbf{k}$ and $\Gamma^\chi_\mathbf{k}$ are the unknown functions to be evaluated, and all their derivatives with respect to energy are taken at the Fermi surface in the limit $T\rightarrow 0$. 

The right-hand side in Eq.~\ref{MB_equation}, i.e., collision integral term incorporates both chirality-breaking and chirality-preserving scattering and is expressed as:
\begin{align}
 I_{coll}[f^{\chi}_{\mathbf{k}}]=\sum_{\chi' \mathbf{k}'}{\mathbf{W}^{\chi \chi'}_{\mathbf{k k'}}}{(f^{\chi'}_{\mathbf{k'}}-f^{\chi}_{\mathbf{k}})},
 \label{Collision_integral}
\end{align}
where, the scattering rate ${\mathbf{W}^{\chi \chi'}_{\mathbf{k k'}}}$ calculated using Fermi's golden rule:
\begin{align}
\mathbf{W}^{\chi \chi'}_{\mathbf{k k'}} = \frac{2\pi n}{\mathcal{V}}|\bra{u^{\chi'}(\mathbf{k'})}U^{\chi \chi'}_{\mathbf{k k'}}\ket{u^{\chi}(\mathbf{k})}|^2\times\delta(\epsilon^{\chi'}(\mathbf{k'})-\epsilon_F).\nonumber\\
\label{Fermi_gilden_rule}
\end{align}
In the above expression \lq n\rq~is impurity concentration, \lq $\mathcal{V}$\rq~is system volume, $\ket{u^{\chi}(\mathbf{k})}$ is chiral spinor, $U^{\chi \chi'}_{\mathbf{k k'}}$ is scattering potential profile, and $\epsilon_F$ is the Fermi energy. We choose $U^{\chi \chi'}_{\mathbf{k k'}}= I_{2\times2}U^{\chi \chi'}$ for elastic impurities, where, $U^{\chi \chi'}$ distinguishes chirality-breaking and chirality-preserving scatterings, which can be controlled in our formalism. We denote the relative magnitude of chirality-breaking to chirality-preserving scattering by the ratio $\alpha = U^{\chi\chi'\neq\chi}/U^{\chi\chi}$ in our formalism. In the context of WSMs, $\alpha$ denotes the ratio of internode to intranode scattering strength, while for SOC-NCMs it denotes the ratio of interband to intraband scattering strength.

In the presence of electric ($\mathbf{E}$) and magnetic ($\mathbf{B}$) fields, semiclassical dynamics of the quasiparticles are modified and governed by the following equation~\cite{son2012berry,knoll2020negative}:
\begin{align}
\dot{\mathbf{r}}^\chi &= \mathcal{D}^\chi_\mathbf{k} \left( \frac{e}{\hbar}(\mathbf{E}\times \mathbf{v}^\chi) + \frac{e}{\hbar}(\mathbf{v}^\chi\cdot \boldsymbol{\Omega}^\chi) \mathbf{B} + \mathbf{v}_\mathbf{k}^\chi\right) \nonumber\\
\dot{\mathbf{p}}^\chi &= -e \mathcal{D}^\chi_\mathbf{k} \left( \mathbf{E} + \mathbf{v}_\mathbf{k}^\chi \times \mathbf{B} + \frac{e}{\hbar} (\mathbf{E}\cdot\mathbf{B}) \boldsymbol{\Omega}^\chi \right),
\label{Couplled_equation}
\end{align}
where, $\mathbf{v}_\mathbf{k}^\chi = ({\hbar}^{-1}){\partial\epsilon^{\chi}(\mathbf{k})}/{\partial\mathbf{k}}$ is band velocity, $\boldsymbol{\Omega}^\chi_\mathbf{k} = -\chi \mathbf{k} /2k^3$ is the Berry curvature, and $\mathcal{D}^\chi_\mathbf{k} = (1+e\mathbf{B}\cdot\boldsymbol{\Omega}^\chi_\mathbf{k}/\hbar)^{-1}$ is factor by which density of states is modified due to presence of the Berry curvature. Self-rotation of the Bloch wave packet also gives rise to an orbital magnetic moment (OMM) $\mathbf{m}^\chi_\mathbf{k}$~\cite{xiao2010berry}.  We rotate the magnetic field along the $xz-$plane:  $\mathbf{B} = B (\cos{\gamma},0,\sin{\gamma})$, i.e., for $\gamma=\pi/2$ both the fields are parallel to each other. 

The focus of this work is to investigate the effect of chiral-anomaly term, and we therefore neglect the Lorentz force term. This also allows us to make analytical progress. We point out that this approximation becomes exact in the limit $\gamma\rightarrow\pi/2$. Even if $\gamma<\pi/2$, the Lorentz force magnitude is comparatively smaller~\cite{ahmad2021longitudinal}.
Keeping terms up to the second order in the electric field, the Boltzmann transport equation reduces to the following set of equations:
\begin{align}
&\mathcal{D}^{\chi}_\mathbf{k}\left[{v^{\chi,z}_{\mathbf{k}}}+\frac{eB\sin{\gamma}}{\hbar}(\mathbf{v^{\chi}_k}\cdot\mathbf{\Omega}^{\chi}_k)\right]
 = \sum_{\chi' \mathbf{k}'}{\mathbf{W}^{\chi \chi'}_{\mathbf{k k'}}}{(\Lambda^{\chi'}_{\mathbf{k'}}-\Lambda^{\chi}_{\mathbf{k}})}.\label{Eq_boltz_E_1}
 \\
&\mathcal{D}^{\chi}_\mathbf{k}\frac{\partial}{\partial \epsilon^\chi_\mathbf{k}} \left(\frac{\partial f_0}{\partial \epsilon^\chi_\mathbf{k}} ~\Lambda^{\chi}_{\mathbf{k}}\right)
\left[{v^{\chi,z}_{\mathbf{k}}}+\frac{eB\sin{\gamma}}{\hbar}(\mathbf{v^{\chi}_k}\cdot\mathbf{\Omega}^{\chi}_k)\right]=\nonumber\\
& \sum_{\chi' \mathbf{k}'}{\mathbf{W}^{\chi \chi'}_{\mathbf{k k'}}}\left(\Gamma^{\chi'}_{\mathbf{k}'}\frac{\partial}{\partial \epsilon^{\chi'}_\mathbf{k'}} \left(\frac{\partial f_0}{\partial \epsilon^{\chi'}_{\mathbf{k}'}}\Lambda^{\chi'}_{\mathbf{k}'}\right)-\Gamma^{\chi}_{\mathbf{k}}\frac{\partial}{\partial \epsilon^{\chi}_\mathbf{k}} \left(\frac{\partial f_0}{\partial \epsilon^{\chi}_{\mathbf{k}}}\Lambda^{\chi}_{\mathbf{k}}\right)\right)
\label{Eq_boltz_E_2}
\end{align}

Eq.~\ref{Eq_boltz_E_1} can be solved for $\Lambda^\chi$, which can then be used to solve for $\Gamma^\chi$ in Eq.~\ref{Eq_boltz_E_2}, and then the distribution function is evaluated using Eq.~\ref{Eq:g1}. Once the distribution function is evaluated, the current density can be evaluated as:
\begin{align}
    \mathbf{J}=-e\sum_{\chi,\mathbf{k}} f^{\chi}_{\mathbf{k}} \dot{\mathbf{r}}^{\chi}.
    \label{Eq:J_formula}
\end{align}
We primarily focus on the second-order anomalous Hall response induced by the chiral anomaly, which is given by 
\begin{align}
    \mathbf{J}^\mathrm{CNLAH}=-\frac{e^2}{\hbar}\sum_{\chi,\mathbf{k}} \mathcal{D}^\chi_\mathbf{k} g^{\chi}_{\mathbf{k}}  (\mathbf{E}\times \mathbf{v}^\chi_\mathbf{k})
    \label{Eq:CNLH_formula}
\end{align}
To evaluate all the different responses, $\mathbf{J}^{\mathrm{CNLAH}}$ is written as:
\begin{align}
    J_{\alpha} = \sum_{\beta, \chi} \sigma^{\chi}_{\alpha \beta} E_{\beta}E_{\beta},
    \label{Eq:CNLH_componetns_formula}
\end{align}
with, $\alpha, \beta = x,y,z$. Comparison of Eq.~\ref{Eq:CNLH_formula} and Eq.~\ref{Eq:CNLH_componetns_formula} gives different components of nonlinear conductivity tensor ($\sigma^{\chi}_{\alpha \beta}$). For, $\mathbf{E} =E \hat{z}$, the anomalous velocity ($\mathbf{v}^{\chi}_{\mathrm{anom}} \sim \mathbf{E} \times \mathbf{\Omega}^{\chi}_{\mathbf{k}}$) has components in $xy$-plane. Since we rotate the magnetic field in $xz$-plane, we measure the Hall response along the $y$-direction, i.e., we evaluate $\sigma_{zy}$. A component of the nonlinear current is also generated along the $x-$direction ($\sigma_{zx}$), which contributes to the planar nonlinear Hall effect, and is seen to vanish.

Moving on, we define the chiral scattering rate as follows:
\begin{align}
\frac{1}{\tau^{\chi}(\theta,\phi)}=\sum_{\chi'}\mathcal{V}\int\frac{d^3\mathbf{k'}}{(2\pi)^3}(\mathcal{D}^{\chi'}_{\mathbf{k}'})^{-1}\mathbf{W}^{\chi \chi'}_{\mathbf{k k'}}.
\label{Tau_invers}
\end{align}
$\mathbf{W}^{\chi \chi'}_{\mathbf{k k'}}$ is defined in Eq.~\ref{Fermi_gilden_rule} and the corresponding overlap of the Bloch wave function is given by the following expression:
$\mathcal{G}^{\chi\chi'}(\theta,\phi) = [1+\chi\chi'(\cos{\theta}\cos{\theta'} + \sin{\theta}\sin{\theta'}\cos{\phi}\cos{\phi'} + \sin{\theta}\sin{\theta'}\sin{\phi}\sin{\phi'}]$. Note that this expression for $\mathcal{G}^{\chi\chi'}(\theta,\phi)$ holds for both our systems of interest: WSM and SOC-NCM. For chiral particles with a different spinor structure, $\mathcal{G}^{\chi\chi'}(\theta,\phi)$ should be appropriately modified. 
\begin{figure}
    \centering
    \includegraphics[width=\columnwidth]{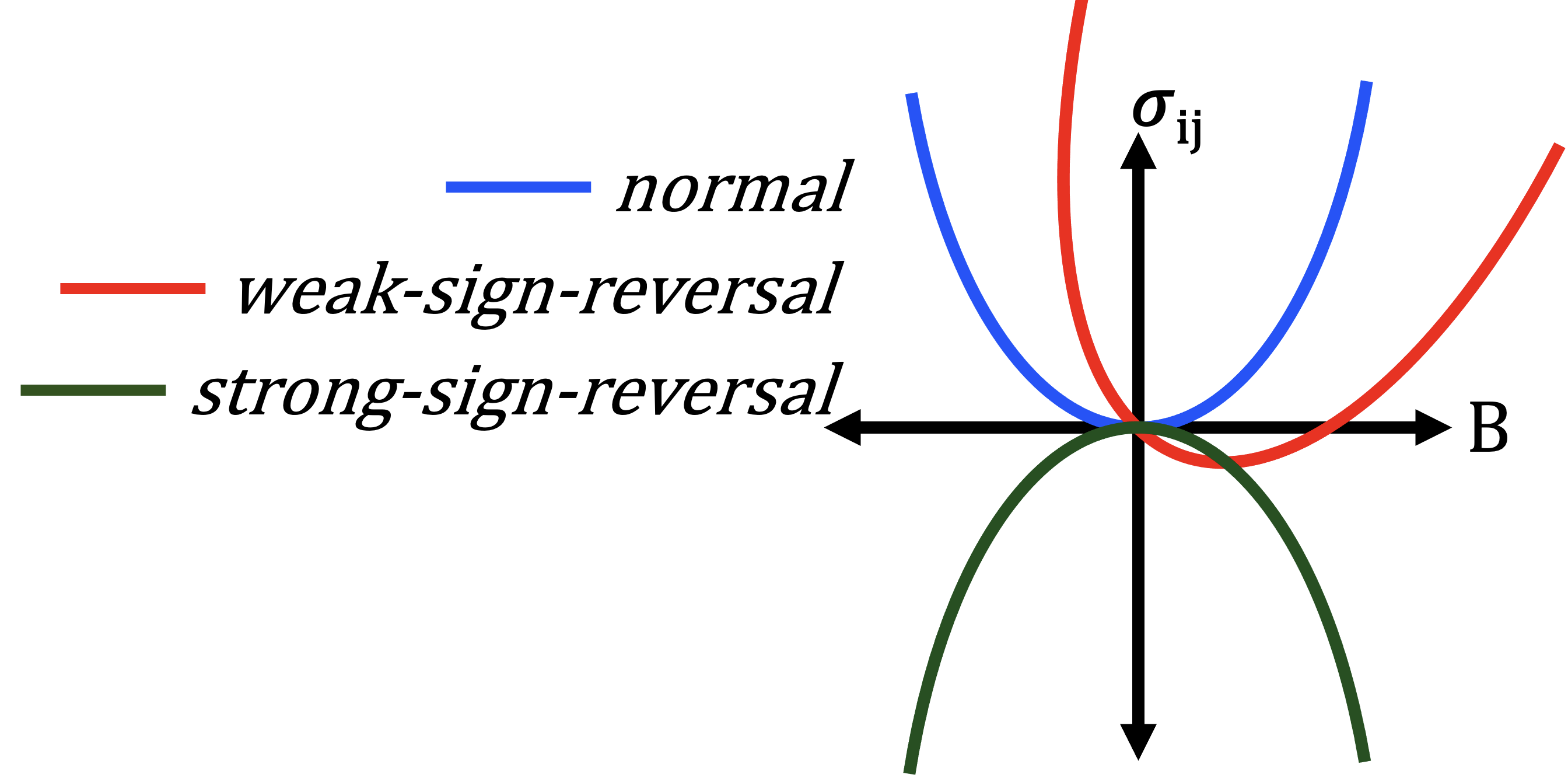}
    \caption{A schematic representation of weak-sign-reversal and strong-sign-reversal of conductivity $\sigma_{ij}$ compared to normal quadratic-in-$B$ conductivity in chiral Weyl systems~\cite{ahmad2023longitudinal,varma2024magnetotransport,ahmad2024geometry}. }
    \label{fig:signreverseschematic}
\end{figure}
Taking Berry phase into account and corresponding change in density of states, $\sum_{k}\longrightarrow \mathcal{V}\int\frac{d^3\mathbf{k}}{(2\pi)^3}\mathcal{D}^\chi_\mathbf{k}$, Eq.~\ref{Eq_boltz_E_1} becomes:
\begin{multline}
l^{\chi}(\theta,\phi) + \frac{\Lambda^{\chi}(\theta,\phi)}{\tau^{\chi}(\theta,\phi)}\\=\sum_{\chi'}\mathcal{V}\int\frac{d^3\mathbf{k}'}{(2\pi)^3} \mathcal{D}^{\chi'}_{\mathbf{k}'}\mathbf{W}^{\chi \chi'}_{\mathbf{k k'}}\Lambda^{\chi'}(\theta',\phi').
\label{MB_in_term_Wkk'}
\end{multline}
Here, $l^{\chi}(\theta,\phi)=\mathcal{D}^{\chi}_{\mathbf{k}}[v^{\chi}_{z,\mathbf{k}}+eB\sin{\gamma}(\mathbf{\Omega}^{\chi}_{k}\cdot \mathbf{v}^{\chi}_{\mathbf{k}})/
\hbar]$ evaluated at the Fermi surface. Eq.~\ref{Tau_invers} and right-hand side of Eq.~\ref{MB_in_term_Wkk'} is reduced to integration over $\theta'$ and $\phi'$,
\begin{align}
\frac{1}{\tau^{\chi}(\theta,\phi)} =  \mathcal{V}\sum_{\chi'} \Pi^{\chi\chi'}\iint\frac{(k')^3\sin{\theta'}}{|\mathbf{v}^{\chi'}_{k'}\cdot{\mathbf{k'}^{\chi'}}|}d\theta'd\phi' \mathcal{G}^{\chi\chi'}(\mathcal{D}^{\chi'}_{\mathbf{k'}})^{-1}.
\label{Tau_inv_int_thet_phi}
\end{align}
\begin{multline}
\mathcal{V}\sum_{\chi'} \Pi^{\chi\chi'}\iint f^{\chi'}(\theta',\phi') \mathcal{G}^{\chi \chi'} d\theta' d\phi'\times[d^{\chi'} - l^{\chi'}(\theta',\phi') \\+ a^{\chi'} \cos\theta' + b^{\chi'} \sin\theta' \cos{\phi'} + c^{\chi'}\sin{\theta'} \cos{\phi'}],
\end{multline}
where, $\Pi^{\chi \chi'} = N|U^{\chi\chi'}|^2 / 4\pi^2 \hbar^2$ and $f^{\chi} (\theta,\phi)=\frac{(k)^3}{|\mathbf{v}^\chi_{\mathbf{k}}\cdot \mathbf{k}^{\chi}|} \sin\theta (\mathcal{D}^\eta_{\mathbf{k}})^{-1} \tau^\chi(\theta,\phi)$. Using the ansatz $\Lambda^{\chi}_{\mathbf{k}}=[d^{\chi}-l^{\chi}(\theta,\phi) + a^{\chi}\cos{\phi} +b^{\chi}\sin{\theta}\cos{\phi}+c^{\chi}\sin{\theta}\sin{\phi}]\tau^{\chi}(\theta,\phi)$, the above equation can be written in following form:
\begin{multline}
d^{\chi}+a^{\chi}\cos{\phi}+b^{\chi}\sin{\theta}\cos{\phi}+c^{\chi} \sin{\theta}\sin{\phi}\\
=\mathcal{V}\sum_{\chi'}\Pi^{\chi\chi'}\iint f^{\chi'}(\theta',\phi')d\theta'd\phi'\\\times[d^{\chi'}-l^{\chi'}(\theta',\phi')+a^{\chi'}\cos{\theta'}+b^{\chi'}\sin{\theta'}\cos{\phi'}+c^{\chi'} \sin{\theta'}\sin{\phi'}].\\
\label{Boltzman_final}
\end{multline}
When this equation is explicitly written, it appears as seven simultaneous equations that must be solved for eight variables. The particle number conservation provides an additional restriction.
\begin{align}
\sum\limits_{\chi}\sum\limits_{\mathbf{k}} g^{\chi}_\mathbf{k} = 0
\label{Eq_sumgk}
\end{align} 
For the eight unknowns ($d^{\pm 1}, a^{\pm 1}, b^{\pm 1}, c^{\pm 1}$), Eq.~\ref{Boltzman_final} and Eq.~\ref{Eq_sumgk} are simultaneously solved with Eq.~\ref{Tau_inv_int_thet_phi}. The nonlinear Hall conductivity induced by chiral anomaly is then evaluated using Eq.~\ref{Eq:CNLH_formula} and Eq.~\ref{Eq:CNLH_componetns_formula}.

\begin{figure*}
    \includegraphics[width=1.95\columnwidth]{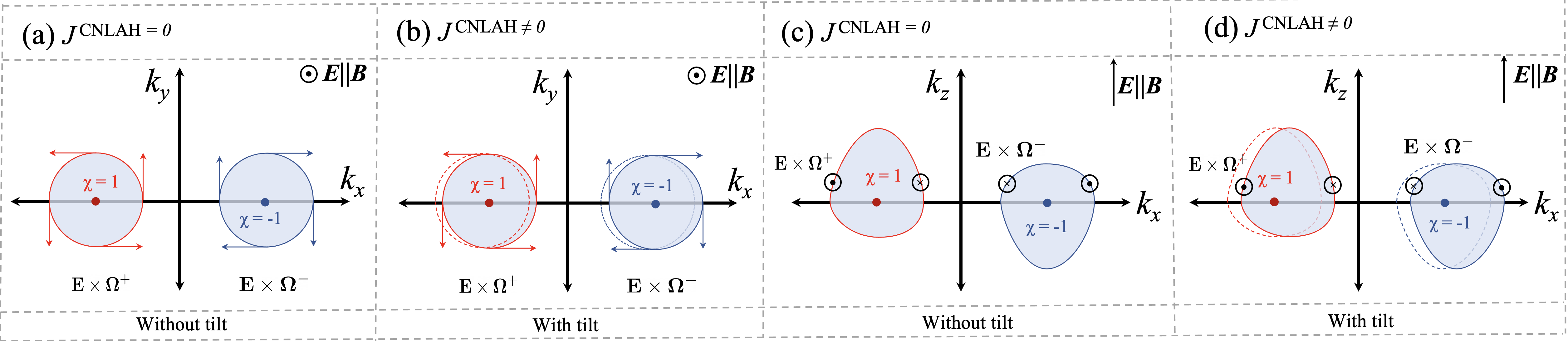}
    \caption{(a) Cross-sectional views of Fermi surface of a WSM at a constant  $k_z=0$ plane in the absence of tilt, and (b) in the presence of tilt. The anomalous velocity vectors $\sim (\mathbf{E}\times \boldsymbol{\Omega}^\chi_\mathbf{k})$ at each $\mathbf{k}$-point on the Fermi surface are indicated by blue and red arrows. (c) Cross-sectional view of Fermi surface of a WSM at a constant $k_y=0$ plane in the absence of tilt. The anomalous velocity vectors now point in and out of the $k_x-k_z$ plane. Due to the effect of orbital magnetic moment, the Fermi surface becomes egg-shaped, and the anomalous velocity vector has a nonuniform magnitude on it. The CNLAH current on each valley is non-zero and opposite in sign and thus the total CNLAH remains zero. (d) Cross-sectional view of Fermi surface of a WSM at a constant $k_y=0$ plane in the presence of tilt. The CNLAH current is of unequal magnitudes on both nodes and therefore the net current does not vanish. In all the plots $\mathbf{E}\parallel\mathbf{B}\parallel \hat{z}$, and the effects of OMM are included.}
\label{fig:WSM_Schematic_Fermi_Surface}
\end{figure*}

\begin{figure*}
    \centering   \includegraphics[width=2\columnwidth]{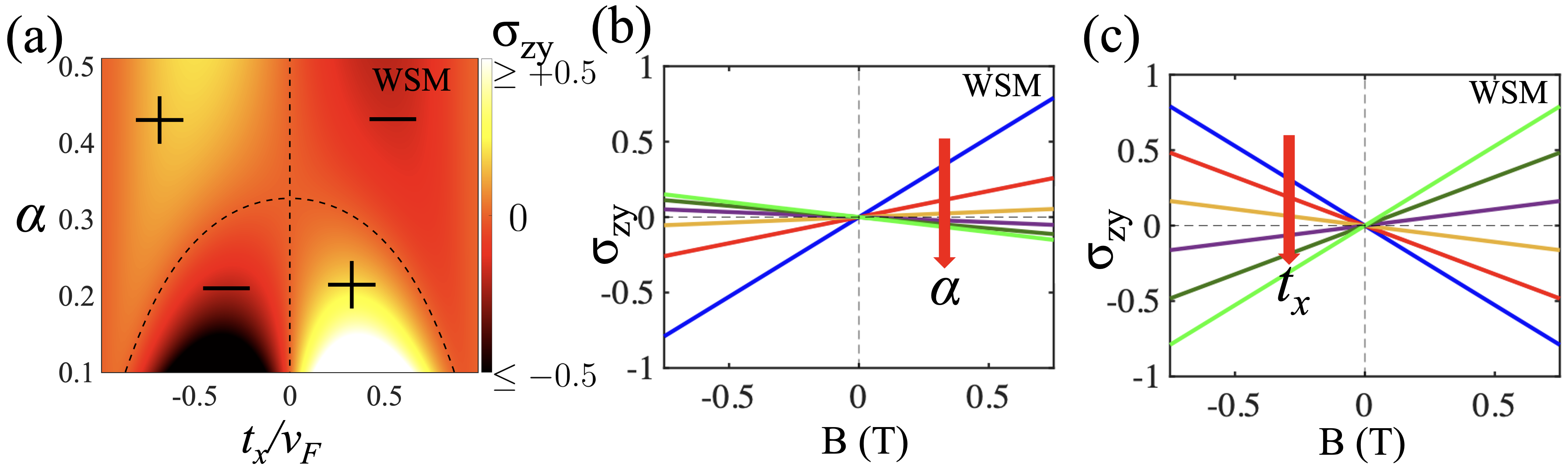}
    \caption{(a) CNLH conductivity $\sigma_{zy}$ as a function of the relative intervalley scattering strength $\alpha$ and the Weyl cone tilt along $x$-direction ($t_x$), for a constant value of magnetic field. Regions of positive and negative magnetoconductivity are separated by black dashed contours and are marked by $+$ and $-$ signs, respectively. (b) $\sigma_{zy}$ as a function of $B$ for different values of $\alpha$ for a constant $t_x$. The red arrow indicates the direction of the increment of $\alpha$, leading to `strong-sign-reversal'. (c) $\sigma_{zy}$ as a function of $B$ for different values of tilt $t_x$ and a constant value of $\alpha$. Here, along the direction of the arrow, we vary tilt $t_x$ from $-0.25~v_F$ to $0.25~v_F$. In all the plots $\gamma=\pi/2$, i.e., the electric and magnetic fields are parallel to each other.}
\label{fig:CNLH_Normal_phase_tx_vary}
\end{figure*}

\begin{figure}
    \includegraphics[width=1\columnwidth]{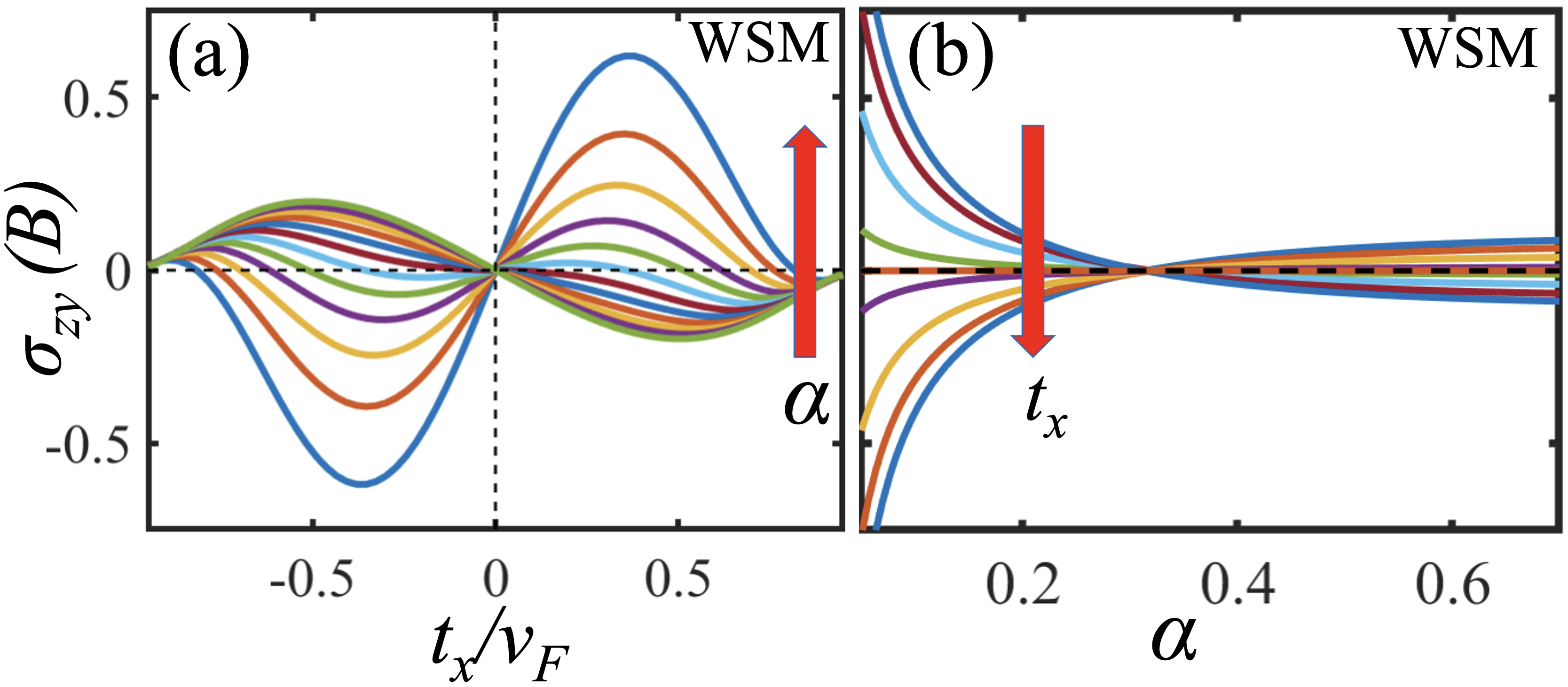}
    \caption{(a) CNLAH conductivity $\sigma_{zy}$ as a function of $t_x$ in WSMs at different values of $\alpha$. The red arrow shows the increment of $\alpha$ from $0.05$ to $1.00$. It is evident that $\sigma_{zy}$ is linear and monotonic only for small values of $\alpha$ and $t_x$. Increasing $\alpha$ leads to sign-reversal of the conductivity. (b) $\sigma_{zy}$ as a function of $\alpha$ for WSMs at different values of $t_{x}$. Here, we have chosen $B\simeq 0.50 T$ and $\gamma=\pi/2$, i.e., the electric and magnetic fields are parallel to each other. The red arrow indicates increment of $t_{x}$ from $-0.25~ v_F$ to $-0.25 ~v_F$. The sweetspot at $\alpha\approx 1/3$, where $\sigma_{zy}\approx0$ is noteworthy. In all the plots $\sigma_{zy}$ has been appropriately normalized.}
\label{fig:CNLH_tx_and_alp_normal_plt}
\end{figure}
\begin{figure*}
    \centering   \includegraphics[width=1.98\columnwidth]{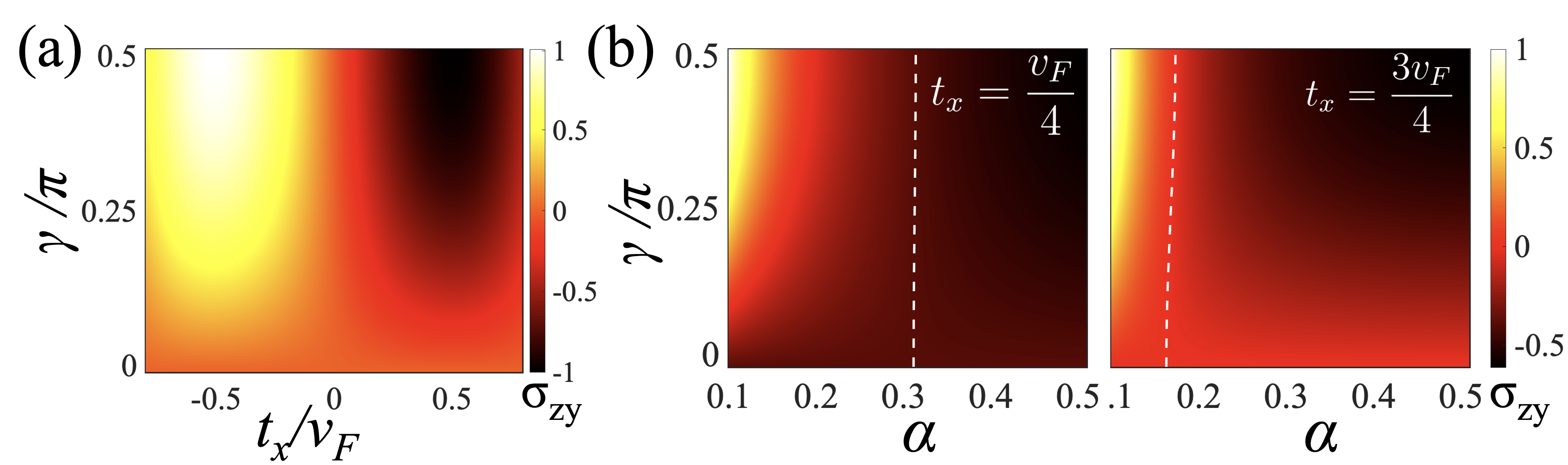}
    \caption{(a) CNLAH conductivity as a function of $t_x$ and $\gamma$ for a constant value of $\alpha$. (b) CNLAH conductivity as a function of $\gamma$ and $\alpha$ for two different values of the tilt parameter $t_x$. The white dashed contour separates the region of positive and negative conductivity. We note that the zero-conductivity contour shows weak dependence on the angle of the magnetic field and a stronger dependence on $t_x$.}
\label{fig:CNLH_Normal_phase_gm_vary}
\end{figure*}

\section{Chiral nonlinear anomalous Hall effect in WSMs}
\subsection{Low-energy Hamiltonian}
We begin with the following low-energy Hamiltonian of a Weyl semimetal:
\begin{multline}
H_{\mathrm{WSM}}(\mathbf{k})=
\sum_{\chi=\pm1} \chi\hbar v_{F}\mathbf{k}\cdot\boldsymbol{\sigma} + \hbar v_{F}(t_z^{\chi} k_z + t_x^{\chi} k_x )I_{2\times2},
\label{Hamiltonian}
\end{multline}
where $\chi$ is the chirality of the Weyl node, $\hbar$ is the reduced Plank constant, $v_{F}$ is the Fermi velocity, $\mathbf{k}$ is the wave vector measured from the Weyl node, $\boldsymbol{\sigma}$ is vector of Pauli matrices, and $t_{x,z}$ are the tilt parameters. The energy dispersion is given by:
\begin{align}
\epsilon^{\chi}_k=\pm{\hbar v_{F}}|k| + \hbar v_{F}(t^{\chi}_z k_z+t^{\chi}_x k_x).
\label{Dispersion}
\end{align}
The constant energy Fermi contour, which is the locus of all points with constant energy $\epsilon$, can be then evaluated to be
\begin{align} k^{\chi} = \frac{\epsilon+\sqrt{\epsilon^{2} - n^{\chi} \chi e v_{F} B \beta_{\theta \phi}}}{n^{\chi}}.
\end{align}
Here, $n^{\chi} = 2 \hbar v_{F} + 2 t_{x} \hbar v_{F} \sin{\theta} \cos{\phi} + 2 t_{z} \hbar v_{F} \cos{\theta}$, and ${\beta_{\theta \phi} = \sin{\theta} \cos{\phi} \cos{\gamma} + \cos{\theta} \sin{\gamma}}$.
The topological nature of following Bloch states of Hamiltonian in Eq.~\ref{Hamiltonian}: $\ket{u^{+}}^{T}=[e^{-i\phi}\cos(\theta /2),\sin(\theta/2)], \ket{u^{-}}^{T}=[-e^{-i\phi}\sin(\theta /2),\cos(\theta/2)]$, gives rise nonzero flux of the Berry curvature $\boldsymbol{\Omega}^{\chi}_\mathbf{k}=-{\chi \mathbf{k}}/{2k^3}$, and its self-rotation gives the anomalous orbital magnetic moment (OMM) $\mathbf{m}^{\chi}_\mathbf{k}=-{\chi e v_{F} \mathbf{k}}/{2k^2}$. Due to the anomalous orbital magnetic moment, the energy dispersion is modified in the presence of the external magnetic field: $\epsilon^\chi_k \rightarrow \epsilon_k - \mathbf{m}^{\chi}_\mathbf{k} \cdot \mathbf{B}$. This changes the spherical Fermi surface to an egg-shaped Fermi surface as schematically displayed in Fig.~\ref{fig:WSM_Schematic_Fermi_Surface}.
Due to change in the dispersion, the band velocity components are also altered, which we evaluate to be:
\begin{align}
v^{\chi}_x&=v_{F}\frac{k_x}{k}+v_Ft^{\chi}_x\nonumber\\
&+\frac{u^{\chi}_{2}}{k^2} \left(\cos{\gamma}\left(1-\frac{2k^2_x}{k^2}\right)+\sin{\gamma}\left(\frac{-2k_x k_z}{k^2}\right)\right), \nonumber\\
v^{\chi}_y&=v_{F}\frac{k_y}{k}\nonumber\\ 
&+\frac{u^{\chi}_{2}}{k^2}\left(\cos{\gamma}\left(\frac{-2k_x k_y}{k^2}\right)+\sin{\gamma}\left(\frac{-2k_y k_z}{k^2}\right)\right),\nonumber\\
v^{\chi}_z&=v_{F}\frac{k_z}{k}+v_Ft^{\chi}_z\nonumber\\ 
&+\frac{u^{\chi}_{2}}{k^2}\left(\cos{\gamma}\left(\frac{-2k_x k_z}{k^2}\right)+\sin{\gamma}\left(1-\frac{2k^2_z}{k^2}\right)\right),
\label{velocity_components}
\end{align}
with $u^\chi_2={\chi e v_{F} B}/{2 \hbar}$.
\subsection{Weak and strong sign-reversal}
The sign of longitudinal magnetoconductivity in Weyl materials has been intensely investigated in prior literature. While chiral anomaly in untilted Weyl semimetals is predicted to show positive LMC for weak internode scattering, it reverses sign for sufficiently strong internode scattering~\cite{knoll2020negative,sharma2020sign,sharma2023decoupling}. On the other hand, even a small amount of tilting in the Weyl cone can result in negative LMC along a particular direction of the magnetic field even for weak internode scattering. However, the reversal in sign in these two cases is fundamentally quite different, which leads to the classification of `strong-sign-reversal' and `weak-sign-reversal' as defined in Ref.~\cite{ahmad2023longitudinal,varma2024magnetotransport}. We briefly review it here. A general expression for the magnetoconductivity tensor can be written as~\cite{ahmad2023longitudinal}
\begin{align}
\sigma_{ij}(B)= \sigma_{ij}^{(2)}+ \sigma_{ij}^{(0)}  (B-B_0)^2,
\label{Eq-sij-fit}
\end{align}
which incorporates (i) normal quadratic $B-$dependence, (ii) linear-in-$B$ dependence and sign change along a particular direction of the magnetic field, and (iii) quadratic-in-$B$ dependence with negative sign, in a single framework.
The features characterizing `weak-sign-reversal' include (i) $B_0 \neq 0$, (ii) $\sigma_{ij}^{(0)} \neq \sigma_{ij}(B=0)$, and (iii) $\mathrm{sign }\; \sigma_{ij}^{(2)}>0$. In this case, the vertex of the magnetoconductivity parabola is shifted from the origin, and the conductivity is of different signs for small positive and negative magnetic fields. However, the orientation of the parabola is still positive, i.e., $\mathrm{sign }\; (\sigma_{ij}^{(2)})>0$.
`Strong-sign-reversal' is characterized by $\mathrm{sign }\; (\sigma_{ij}^{(2)})<0$, which implies a complete reversal of the orientation of the parabola. Tilting of Weyl cones can result in `weak-sign-reversal' while intervalley scattering or strain is generally expected to result in `strong-sign-reversal'~\cite{ahmad2023longitudinal,varma2024magnetotransport}. Fig.~\ref{fig:signreverseschematic} schematically explains the distinction between the two cases. 

\begin{figure}
    \centering
    \includegraphics[width=0.99\columnwidth]{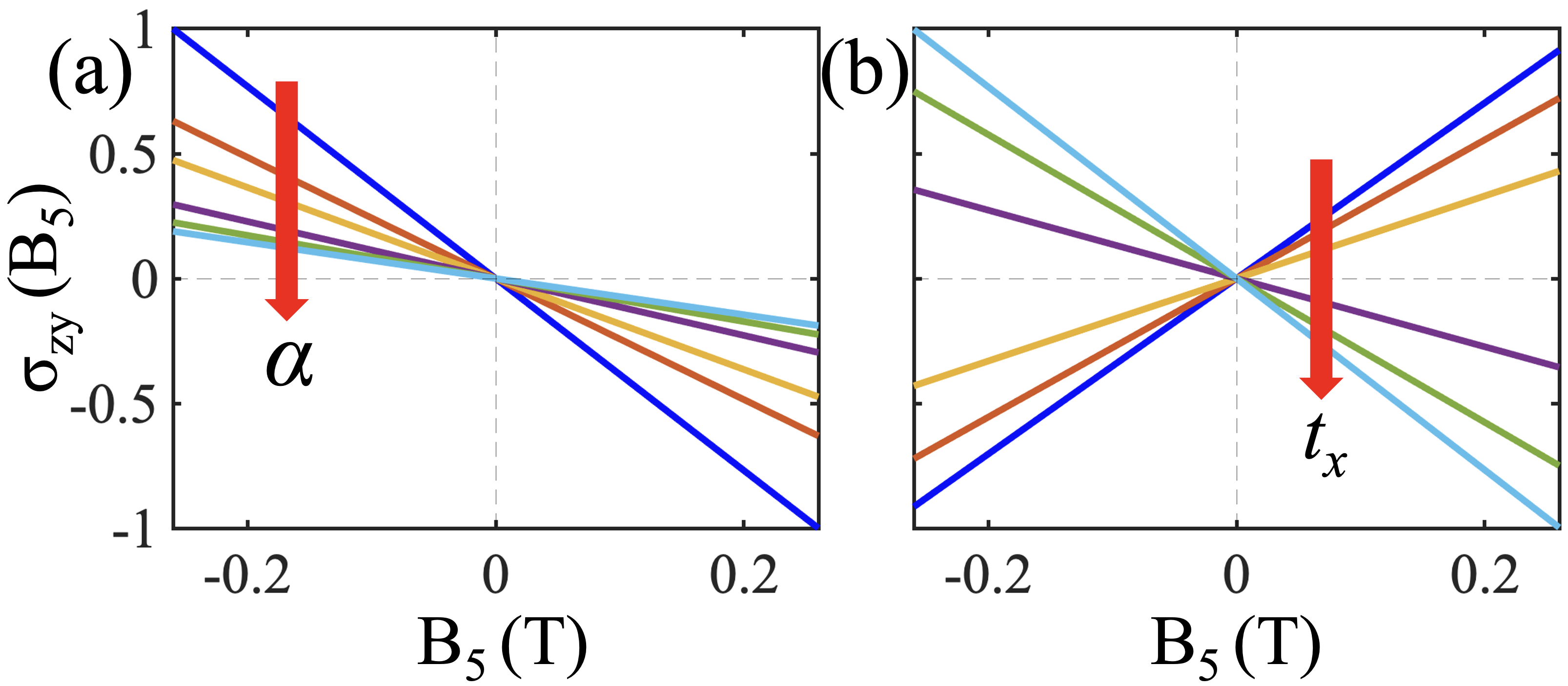}
    \caption{The anomalous nonlinear Hall conductivity as a function of the strain-induced magnetic field $\mathbf{B}_5$. (a) For different intervalley scattering strengths $\alpha$ but fixed $t_x$. (b) For different values of tilt $t_x$ but fixed $\alpha$. The parameters $\alpha$ and $t_x$ increase in the direction of the arrow in both the respective plots. The $\mathbf{B}_5$ field here was chosen parallel to the electric field.}
    \label{fig:wsm_cnlh_B5}
\end{figure}

\begin{figure*}
    \includegraphics[width=1.99\columnwidth]{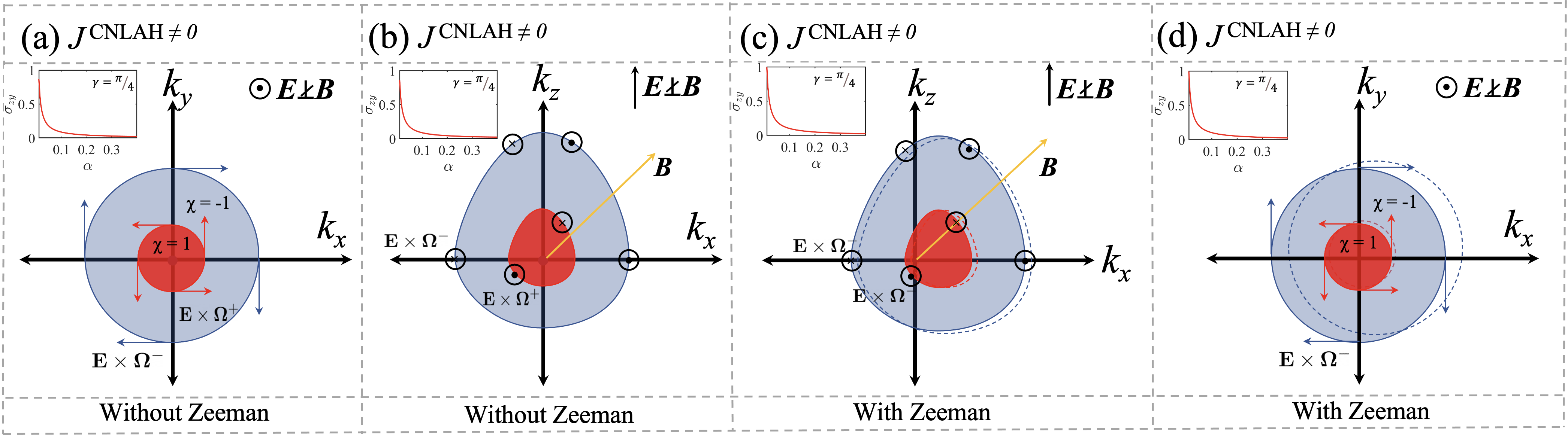}
    \caption{Schematic illustration of the origin of CNLAH in SOC-NCMs. (a) Cross-sectional view of the Fermi surface in the $k_z=0$ plane without the effect of Zeeman coupling. The anomalous velocity at two Fermi surfaces having opposite chirality is marked by blue and red arrows. (b) Cross-sectional view in the $k_y=0$ plane. The anomalous velocity vectors now point in and out of the $k_z-k_x$ plane. (c) Cross-sectional view of the Fermi surface in the $k_y=0$ plane and (d) in the $k_z=0$ plane, including the effect of the Zeeman field that effectively acts like a tilt term in the Hamiltonian.  In all the plots $\mathbf{E}\parallel \hat{z}$, $\mathbf{B} = B(\cos\gamma,0,\sin\gamma)$, and the effect of the OMM has been considered. The insets show the relative magnitudes of the total CNLAH current with respect to the interband scattering parameter $\alpha$. The current is seen to enhance including the effect of  Zeeman coupling.}
\label{fig:SOC_Schematic_Fermi_Surface}
\end{figure*}

\subsection{Nonlinear anomalous Hall conductivity}
We are now in a position to discuss the chiral anomaly-induced nonlinear anomalous Hall conductivity in WSMs.
In the absence of any Weyl cone tilting and effects of orbital magnetic moment, the net anomalous velocity vector vanishes for each node resulting in zero nonlinear anomalous Hall conductivity. When the Weyl cones are untilted and when the effect of orbital magnetic moment is excluded, from symmetry considerations it is easy to conclude that the net CNLAH current vanishes at each node. 
When effects of the orbital magnetic moment are included, the Fermi surface becomes egg-shaped (see Fig.~\ref{fig:WSM_Schematic_Fermi_Surface}) and the net anomalous velocity vector does not vanish, resulting in a nonzero CNLAH current at each node. However, the net current vanishes when the contribution from both nodes is added up because the current at both nodes is of equal magnitudes and opposite signs. Now, when the Weyl cones are further tilted, the nonlinear Hall current at both nodes is unequal in magnitude, resulting in a net non-zero CNLAH current. Fig.~\ref{fig:WSM_Schematic_Fermi_Surface} schematically presents cross-sectional views of the Fermi surface of a Weyl semimetal, highlighting the mechanism resulting in a nonvanishing CNLAH current.

In Fig.~\ref{fig:CNLH_Normal_phase_tx_vary}, we plot the CNLAH conductivity $\sigma_{zy}$ as function of $t_x$ and $\alpha$ for $\gamma=\pi/2$ (parallel electric and magnetic fields). We first note that the CNLAH conductivity is an odd function of tilt $t_x$, in accordance with the findings of Ref.~\cite{li2021nonlinear}. However, we find that CNLAH conductivity is non-monotonic as a function of tilt $t_x$, which is in striking contrast to Ref.~\cite{li2021nonlinear} that reports monotonic behavior with respect to $t_x$. 
We find that even small intervalley scattering results in non-monotonicity. The CNLAH conductivity first increases as a function of $t_x$ and then decreases after reaching a maximum. Furthermore, for a fixed value of $t_{x}$, increasing the internode scattering strength $\alpha$, the magnitude of the CNLAH conductivity decreases and eventually flips sign after a critical value $\alpha_c$, i.e. displays `strong-sign-reversal'. These twin effects cause a prominent `half-lung' like pattern as shown in Fig.~\ref{fig:CNLH_Normal_phase_tx_vary}(a). In Fig.~\ref{fig:CNLH_Normal_phase_tx_vary}(b), we plot CNLAH conductivity as a function of $B$ at different values of the internode scattering strength $\alpha$ for fixed tilt. It is clear that increasing $\alpha$ results in sign-reversal of $\sigma_{zy}$. Now, fixing $\alpha$ and increasing the amount of tilt, $\sigma_{zy}$ also changes sign as shown in Fig.~\ref{fig:CNLH_Normal_phase_tx_vary}(c). The non-monotonicity of the CNLAH effect and the existence of `strong-sign-reversal' are the prominent features we discover, which have been unreported so far. We attribute these to the effects of chirality-violating scattering and global charge conservation that have not been correctly accounted for in earlier studies.  

To gain further insight, we plot $\sigma_{zy}$ as a function of $t_x$ in Fig.~\ref{fig:CNLH_tx_and_alp_normal_plt} (a) for different values of the internode scattering strength $\alpha$. The conductivity is highly non-monotonic--it first increases as a function of $t_x$ and decreases and eventually becomes close to zero when $t_x\approx 1$. Interestingly, we discover that as $\alpha$ is increased, (i) the conductivity increases, (ii) then quickly falls to zero for some value of $t_x<1$, (iii) then becomes negative, and (iv) finally approaches zero again when $t_x\approx 1$. When $\alpha$ is increased, $\sigma_{zy}$ falls to zero and becomes negative at smaller and smaller values of $t_x$. When $\alpha$ is large enough, the conductivity $\sigma_{zy}$ eventually reverses sign at $t_x\approx 0$.
In Fig.~\ref{fig:CNLH_tx_and_alp_normal_plt} (b), we plot $\sigma_{zy}$ as a function of $\alpha$ for different values of the $t_x$. Remarkably, we find a sweet-spot at $\alpha\approx 1/3$, where $\sigma_{zy}\approx 0$ for all values of $t_x\lesssim 0.25 v_F$. 

Having discussed the CNLAH conductivity for collinear electric and magnetic fields and the effect of $\alpha$, we now discuss the case when $\mathbf{E}$ and $\mathbf{B}$ are noncollinear, since in many experimental setups, the effect of rotating the magnetic field is investigated. In Fig.~\ref{fig:CNLH_Normal_phase_gm_vary}(a) we plot $\sigma_{zy}$ as a function of tilt $t_{x}$ and $\gamma$ for a finite  value of $\alpha$. The conductivity is an odd function of $t_x$, and is a non-monotonic function of the tilt for all values of $\gamma$. In Fig.~\ref{fig:CNLH_Normal_phase_gm_vary}(b), we plot $\sigma_{zy}$ as a function of tilt $\alpha$ and $\gamma$ for two different fixed values of  value of $t_x$. For all angles of the magnetic field $\gamma$, the conductivity shows strong-sign reversal as a function of the intervalley scattering strength $\alpha$, however, the dependence of the zero-conductivity contour (separating the regions of positive and negative conductivity) is seen to be weak, unlike linear longitudinal magnetoconductivity that shows a stronger dependence on $\gamma$~\cite{ahmad2021longitudinal}. The dependence of the zero-conductivity contour is stronger on $t_x$, as we also note from Fig.~\ref{fig:CNLH_Normal_phase_tx_vary}. 
\subsection{Effects of strain}
We next discuss the effect of strain on the nonlinear anomalous Hall conductivity. In a topological protected Weyl semimetal, Weyl nodes are separated in the momentum space by a finite vector $\mathbf{b}$. The vector $\mathbf{b}$ is also interpreted as an axial gauge field because of its opposite coupling to Weyl nodes of opposing chiralities~\cite{goswami2013axionic,volovik1999induced,liu2013chiral,grushin2012consequences,zyuzin2012topological}. A position-dependent $\mathbf{b}$ vector generates an axial magnetic field (denoted as $\mathbf{B}_5=\nabla\times \mathbf{b}$), which also couples oppositely to Weyl nodes of opposite chirality. Such a scenario can arise if Weyl semimetals are subjected to an inhomogeneous strain profile. The effective magnetic field experienced by a fermion at node $\chi$ is therefore $\mathbf{B}\longrightarrow\mathbf{B}+\chi\mathbf{B}_5$. Recent works have studied the role of strain in longitudinal and planar Hall conductivity~\cite{grushin2016inhomogeneous,ghosh2020chirality,ahmad2023longitudinal}, however, its role in the nonlinear anomalous Hall conductivity remains unexplored. 

In Fig.~\ref{fig:wsm_cnlh_B5} we plot the nonlinear Hall conductivity $\sigma_{zy}$ as a function of the strain induced $\mathbf{B}_5$ field. As the intervalley scattering strength is increased the conductivity is suppressed. However, unlike Fig.~\ref{fig:CNLH_Normal_phase_tx_vary}, where we examined the effect of an external magnetic field, there is no strong-sign-reversal for large values of the intervalley scattering strength. 
Furthermore, we find that strain-induced nonlinear $\sigma_{zy}$ also changes sign as a function of the tilt parameter as seen in Fig.~\ref{fig:wsm_cnlh_B5} (b). 
The measurement of nonlinear anomalous Hall conductivity in Weyl semimetals (i) in the presence of strain but the absence of magnetic field, and (ii) in the absence of strain but the presence of external magnetic field, can provide us crucial insights into the role and strength of internode scattering. For instance, if the measured nonlinear conductivity is negative in both scenarios, it is strongly suggestive of large internode scattering. Conversely, if the conductivity is positive in one case and negative in the other, it is indicative of weak internode scattering.

 \begin{figure*}
    \centering
    \includegraphics[width=1.99\columnwidth]{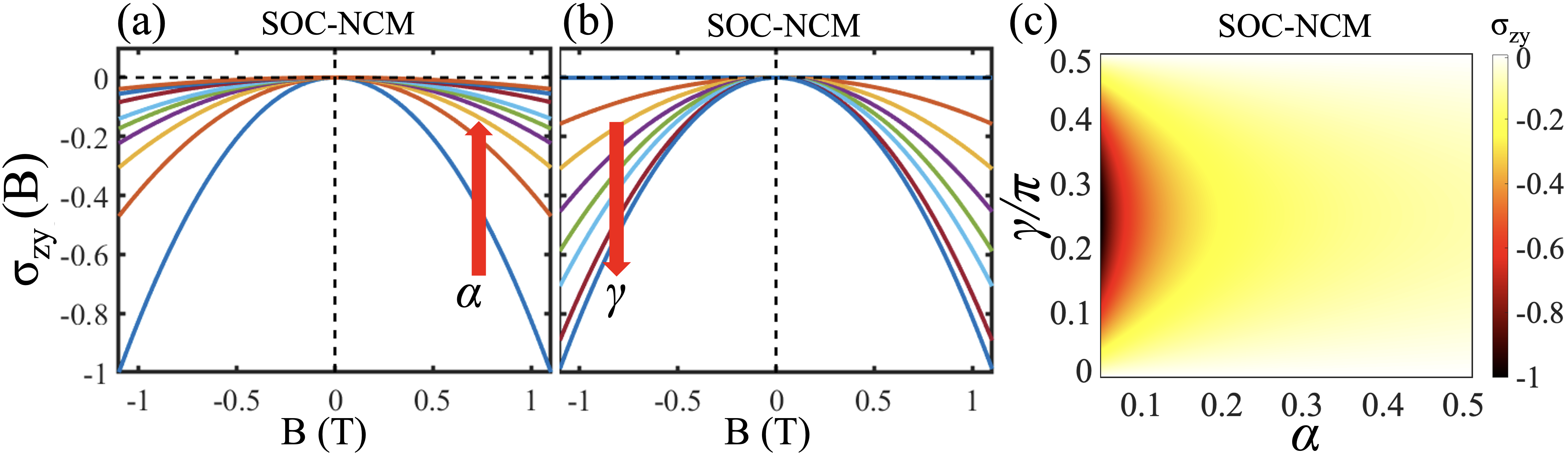}
    \caption{CNLAH conductivity for spin-orbit coupled noncentrosymmetric metals for (a) different values of the interband scattering strength $\alpha$, and (b) for different values of $\gamma$ increasing from zero to $\pi/4$. In both plots, the corresponding parameters increase in the direction of the arrow. (c) The corresponding color plot indicates that the conductivity peaks at $\pi/4$ and decreases with increasing $\alpha$.}
    \label{fig:cnlahe_socnsm}
\end{figure*}

\section{CNLAH in spin-orbit coupled noncentrosymmetric metals}
\subsection{Low-energy Hamiltonian}
We begin with the following low-energy Hamiltonian of a spin-orbit coupled noncentrosymmetric metal expanded near the high-symmetry point~\cite{mukherjee2012order,varma2024magnetotransport}
\begin{align}
    H(\mathbf{k}) = \frac{\hbar^2 k^2}{2m} + \hbar \vartheta \mathbf{k}\cdot\boldsymbol{\sigma} 
    \label{Eq_H_socncm}
\end{align}
were $m$ is the effective mass, $\vartheta$ incorporates the spin-orbit coupling parameter. We couple the system to a Zeeman field given by~\cite{cano2017chiral,sharma2017nernst,gadge2022anomalous} 
\begin{align}
    H_z = -\mathbf{M}\cdot\boldsymbol{\sigma},
\end{align}
where $\mathbf{M}$ is related to the external magnetic field by $\mathbf{M} = -g\mu_B \mathbf{B}/2$, where $\mu_{B}$ is the Bohr magneton and $g$ is Landé g-factor ($g \sim 50$ \cite{xie2021kramers,yakunin2010spin}). The resultant Hamiltonian, including the Zeeman field, becomes 
\begin{align}
    H(\mathbf{k}) = \frac{\hbar^2 k^2}{2m} + \hbar \vartheta \left(\mathbf{k}+\frac{\mathbf{M}}{\hbar \vartheta}\right)\cdot\boldsymbol{\sigma},
\end{align} 
A change of variables ($\mathbf{k}\rightarrow \mathbf{k}-\mathbf{M}/\hbar\vartheta$) yields 
\begin{align}
    H(\mathbf{k}) = \frac{\hbar^2k^2}{2m} + \hbar\vartheta\mathbf{k}\cdot\boldsymbol{\sigma} + \hbar\vartheta(k_x t_x + k_zt_z) + E_0,
\end{align}
where $E_0 = M^2/2m\vartheta^2$ is an irrelevant constant energy shift, $t_x = -M_x/m\vartheta^2$, $t_z = -M_z/m\vartheta^2$. Remarkably, in the new reference frame, the Hamiltonian resembles that of a tilted Weyl semimetal! The effect of the Zeeman coupling is therefore to tilt the Fermi surfaces just like the tilted Weyl cones of a Weyl semimetal. Note that the effective tilt is proportional to the amount of Zeeman coupling and inversely proportional to the effective mass $m$. Therefore, for a purely relativistic $\mathbf{k}\cdot\boldsymbol{\sigma}$ Hamiltonian, where the effective mass term $\sim k^2/m\rightarrow 0$, the effective tilt term vanishes. Hence this property of the noncentrosymmetric metal is distinct from an inversion asymmetric Weyl semimetal where the effective mass term $\sim k^2/m\rightarrow 0$ is absent. 
The energy spectrum is evaluated to be:
\begin{align}
    \epsilon_\mathbf{k}^\lambda = \frac{\hbar^2 k^2}{2 m} + \lambda \hbar \vartheta {k}
    +\hbar \vartheta (k_{x} t_{x} + k_{z} t_{z}) + E_{\mathrm{0}},
\label{eq:E_Eigen_value_SOC_3_in_q}
\end{align}
with, $\lambda=\pm1$ representing two spin-orbit split bands. We note that both the Fermi surfaces are tilted along the same direction as a result of the Zeeman field. To obtain the constant energy Fermi contour, we need to add the orbital magnetic moment coupling to the energy spectrum and invert Eq.~\ref{eq:E_Eigen_value_SOC_3_in_q}. This yields a cubic equation in $k$ that needs to be solved for $k=k(\theta,\phi)$. Since the analytical expression is lengthy and uninteresting, we do not provide it here. The change of variables is implemented straightforwardly in the Boltzmann equation. The Jacobian remains invariant, but the constant energy Fermi contour is appropriately modified while integrating over a constant energy surface in the Boltzmann equation.

Without loss of generality, we assume that the chemical potential lies above the nodal point $\mathbf{k}=0$, and hence the Fermi surface is composed of two disjointed surfaces as shown in Fig.~\ref{fig:SOC_Schematic_Fermi_Surface}. Both the surfaces enclose a nontrivial flux of Berry curvature, which is of the same magnitude but opposite sign. This is similar to the case of a Weyl semimetal where the Berry curvature is of the same magnitude but opposite signs at the two valleys. Interestingly, in SOC-NCM, the anomalous orbital magnetic moment 
($\mathbf{m}^\lambda_\mathbf{k}$) has the same sign and magnitude, which is different from a Weyl semimetal where the signs are reversed at the two nodes. With the application of an external magnetic field, the orbital magnetic moment couples to it as $-\mathbf{m}^\lambda_\mathbf{k}\cdot\mathbf{B}$, leading to the oval-shaped Fermi surfaces as shown in Fig.~\ref{fig:SOC_Schematic_Fermi_Surface}. In Weyl semimetal, the coupling is opposite in the two valleys, and thus, the shapes of Fermi the surfaces are reversed (see Fig.~\ref{fig:WSM_Schematic_Fermi_Surface}).

\subsection{Nonlinear anomalous Hall conductivity}
In Fig.~\ref{fig:SOC_Schematic_Fermi_Surface} (a) and (b), we plot a cross-sectional view of the Fermi surface of the SOC-NCM including the effect of the orbital magnetic moment but without considering Zeeman coupling (which is the effective tilt term). The net anomalous velocity vector ($\sim \mathbf{E}\times\boldsymbol{\Omega}_\mathbf{k}$) at both the nodes does not vanish since the Fermi surface is no longer symmetrical around the $k_x-k_y$ plane. Furthermore, the magnitudes of the CNLAH current at the two nodes are not of equal magnitudes (unlike the case of a WSM). This results in a net nonzero CNLAH current, unlike the WSM where the orbital magnetic moment alone does not result in a nonzero current. Fig.~\ref{fig:SOC_Schematic_Fermi_Surface} (c) and (d) depict the effect of including Zeeman coupling, which further introduces an asymmetry in the Fermi surface and enhances the total CNLAH current. 

In Fig.~\ref{fig:cnlahe_socnsm}, we plot the nonlinear anomalous Hall conductivity for spin-orbit noncentrosymmetric metal described in Eq.~\ref{Eq_H_socncm}. We find that the nonlinear conductivity $\sigma_{zy}$ is quadratic in the magnetic field, in contrast to a Weyl semimetal where $\sigma_{zy}$ is seen to be linear in $B$. An additional $B-$dependence enters in SOC-NCMs because (i) the current here is driven by anomalous orbital magnetic moment unlike in WSM where it is driven by a finite tilt, and (ii) the generated effective tilt due to the Zeeman coupling is field-dependent, unlike in WSMs, where a constant tilt that is inherent to the bandstructure is assumed.  
Furthermore, in SOC-NCMs, we find the conductivity $\sigma_{zy}$ to be negative, which is suppressed with increasing intervalley scattering strength, but importantly does not flip its sign. This is contrasted to the case of Weyl semimetal where a strong-sign-reversal is observed. This crucial difference is attributed to the different nature of the orbital magnetic moment in both cases. 
The behavior of $\sigma_{zy}$ with the angle of the magnetic field $\gamma$ in the current case is also of special interest. Unlike the case of Weyl semimetal, where $\sigma_{zy}$ is maximum when $\gamma=\pi/2$ (when the electric and magnetic field are parallel to each other), in SOC-NCMs, the conductivity is maximum when $\gamma=\pi/4$, i.e., the electric and magnetic field are at $\pi/4$ angle with respect to each other.
This important distinction is understood as follows. First, we observe that in WSMs, the current is driven by a finite tilt of the Weyl cones along the $k_x$ direction. Therefore the current is maximum when the magnetic field points along $\hat{z}-$direction, parallel to the electric field. In SOC-NCMs, the current is driven by anomalous magnetic moment. 
Now, when $\mathbf{E}\parallel\mathbf{B}\parallel \hat{z}$, integral of the anomalous velocity vector $({v}_\textrm{anom}\propto \Omega_x\propto\cos\phi)$ vanishes due to azimuthal symmetry, and therefore the net anomalous current vanishes as well. In WSMs, azimuthal symmetry is destroyed due to a finite $t_x$, even when $\mathbf{E}\parallel\mathbf{B}\parallel \hat{z}$.


\section{Conclusions}
In this work, we advance the theoretical understanding of the chiral anomaly-induced nonlinear anomalous Hall effect (CNLAHE) in three-dimensional chiral fermionic systems, with a particular focus on Weyl semimetals (WSMs) and spin-orbit coupled noncentrosymmetric metals (SOC-NCMs). By rigorously incorporating momentum-dependent chirality-preserving and chirality-breaking scattering processes, as well as global charge conservation, we address critical gaps in the existing models, thereby providing a more robust and comprehensive framework for analyzing CNLAHE.
In the context of Weyl semimetals, we uncover a complex, nonmonotonic relationship between the nonlinear anomalous Hall conductivity and the Weyl cone tilt. This behavior is notably sensitive to the strength of internode scattering, leading to a `strong-sign-reversal' of the conductivity. Moreover, we also investigate the effects of strain-induced chiral gauge fields on CNLAHE, demonstrating that while such strain can indeed generate nonlinear Hall effects, it does so without inducing a sign reversal in the conductivity. Experiments performed with and without external strain in WSMs can shed light on the role of internode scattering by comparing the nonlinear anomalous Hall conductivity (NLAHC) in both scenarios. 
For spin-orbit coupled noncentrosymmetric metals, we reveal that the anomalous orbital magnetic moment is sufficient to drive a large nonlinear conductivity, which is distinguished by its negative sign, regardless of the strength of interband scattering, and its quadratic dependence on the magnetic field. This behavior starkly contrasts with the linear magnetic field dependence observed in WSMs and highlights the fundamental differences between these two classes of materials. We also identify the Zeeman coupling of the magnetic field as a crucial factor that acts as an effective tilt term, further amplifying the CNLAHE in SOC-NCMs.
The theoretical insights presented in this work extend the current understanding of CNLAHE in chiral quasiparticles and provide a critical foundation for current and upcoming experimental investigations. 
\bibliography{biblio.bib}
\end{document}